\documentclass[10pt,twocolumn]{article} 
\usepackage{graphicx} 
\usepackage{dcolumn}
\usepackage{bm}
\usepackage{hyperref}
\pdfoutput=1
\usepackage{amssymb} 
\usepackage{amsmath}
\usepackage{color} 
\usepackage[margin=1in]{geometry}
\usepackage{authblk}

\newcommand{\vect}{{\bf x}}

\begin{document}

\author[1]{J.-B. Delfau\thanks{A.A@university.edu}}
\author[2]{J. Molina\thanks{B.B@university.edu}}
\author[3]{M. Sano\thanks{C.C@university.edu}}
\affil[1]{Institute for Cross-Disciplinary Physics and Complex Systems, University of the Balearic Islands, Palma de Mallorca E-07122 , Spain}
\affil[2]{Department of Chemical Engineering, Kyoto University, Kyoto 615-8510, Japan}
\affil[3]{Department of Physics, The University of Tokyo, 7-3-1 Hongo, Bunkyo-ku, Tokyo 113-0033, Japan}

\title{Collective behavior of strongly confined suspensions of squirmers}

\twocolumn[
  \begin{@twocolumnfalse}
	\maketitle
\begin{abstract}
We run numerical simulations of strongly confined suspensions of model spherical swimmers called ``squirmers''. Because of the confinement, the Stokeslet dipoles generated by the particles are quickly screened and the far-field flow is dominated by the source dipole for all the different kinds of squirmers. However, we show that the collective behavior of the suspension still depends on the self-propelling mechanism of the swimmers as polar states can only be observed for neutral squirmers. We demonstrate that the near-field hydrodynamic interactions play a crucial role in the alignment of the orientation vectors of spherical particles. Moreover, we point out that the enstrophy and the fluid fluctuations of an active suspension also depend on the nature of the squirmers.
\end{abstract}
{\let\newpage\relax\maketitle}
\vspace{1cm}
  \end{@twocolumnfalse}
	]

Understanding the physics of ``active matter'' has become a major preoccupation in statistical physics. Self-propelling objects being fundamentally out-of-equilibrium, they can exhibit various interesting collective behaviors: experimental works pointed out swarming in systems of bacteria\cite{Cisneros07, Drescher11, Sokolov07, Sokolov12, Zhang10, Zhang12} or self-propelled droplets\cite{Thutupalli11}, dynamic clustering\cite{Theurkauff12,Palacci13} and phase separation of Janus particles\cite{Buttinoni13} or directional motion in suspensions of rotating colloids\cite{Bricard13}, vibrated rods\cite{Kumar14} and asymmetric hard-disks\cite{Deseigne10,Deseigne12}. The onset of directional motion in suspensions of swimmers has been intensively studied numerically and theoretically to try to determine in what kind of systems this collective behavior may be observed. Many numerical simulations focused on a simple and well-known model of swimmers called ``squirmers'' introduced for the first time by Lighthill\cite{Lighthill52} and refined further by Blake\cite{Blake71}. It was pointed out that the self-propelling mechanism of these squirmers has a strong influence on the collective dynamics and that only some specific kinds of active particles can exhibit polar states\cite{Ishikawa08, Evans11, Zottl14}. These results were explained by demonstrating that if the motion of the swimmers generates Stokeslet dipoles, these singularities will dominate the far-field hydrodynamic flow and destabilize the polar state\cite{Simha02,Saintillan07,Saintillan08,Baskaran09}. Yet when it comes to strongly confined suspensions, the situation is quite different: in that case, these singularities are screened exponentially\cite{Liron76} and hydrodynamics interactions become independent of the details of the self-propelling mechanism, at least in the far-field approximation\cite{Saintillan08, Baskaran09}. It was thus predicted that the usual distinction between the different kinds of squirmers (pushers, pullers, neutral squirmers) should become irrelevant for strongly confined suspensions\cite{Brotto13}. However, this statement only holds if we assume that the collective dynamics of a suspension is mainly determined by the far-field hydrodynamic interactions. In particular, we can wonder if the near-field interactions do not play a significant role in the alignment of the orientation vectors of the swimmers, which is essential in the formation of polar states. To our knowledge, this question has not been clearly answered yet for squirmers.\\
\begin{figure}[!ht]
\begin{tabular}{cc}
\includegraphics[width=7 cm]{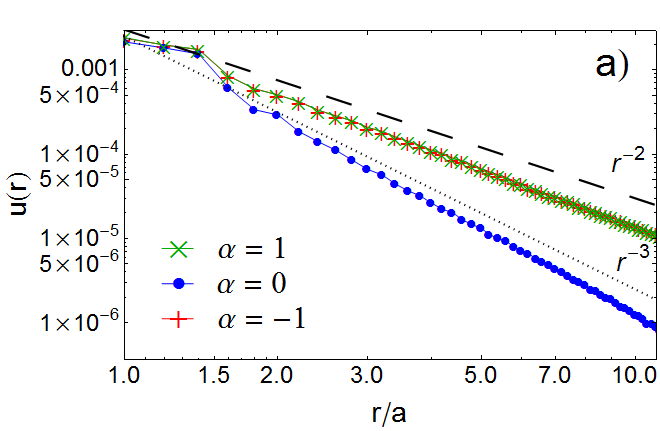} \\
\includegraphics[width=7 cm]{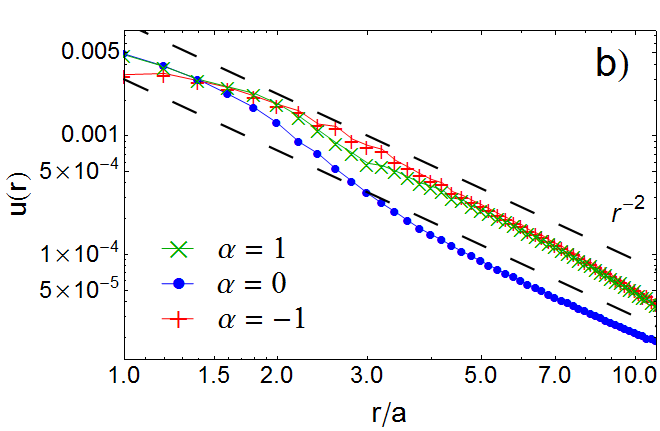}
\end{tabular}
\caption{\label{vfield}Velocity of the flow field $u$ generated by three different kinds of squirmers with respect to the distance $r/a$. a) No confinement. b) Two no-slip walls separated by $h=5.2\, a$. The dashed and dotted black lines correspond to $r^{-2}$ and $r^{-3}$ respectively.}
\end{figure}
In this paper, we run periodic simulations of self-propelled spherical particles with 2D orientations at a constant density. The system is confined vertically by two walls with no-slip boundary conditions. Our numerical simulations are based on the ``Smooth-profile'' method (SPM), which has been described in details in previous publications\cite{Nakayama08}. The key concept of this method is to consider that the interface between the particles and the fluid is not discrete but continuous, with a given thickness. We define a continuous density function $\psi (\vect{r},t)$ such as $\psi =1$ inside the particles, $0 < \psi < 1$ inside the interface and $\psi =0$ in the fluid. If $\vect{u}_p(\vect{r},t)$ is the velocity of the particles, then the function $\psi \vect{u}_p(\vect{r},t)$ is a continuous velocity field for the particles over the whole system. The following modified Navier--Stokes equation is then solved assuming momentum conservation between the fluid and the particles
\begin{equation}
\left\{
\begin{array}{ll}
\partial_t\vect{v} + (\vect{v} \cdot \nabla) \vect{v} & = -\frac{\nabla p}{\rho} + \eta \nabla^2 \vect{v} + \left( \psi \vect{f}_p + \vect{f}_{sq}\right)\\
& \\
\nabla \cdot \vect{v} = 0
\end{array}
\right.
\label{NS_equation}
\end{equation}
where $\vect{v}$ is the total fluid velocity given by $\vect{v} = \psi \vect{u}_p + (1-\vect{u})$ ($\vect{u}$ being the fluid velocity), $p$ is the pressure, $\eta$ the kinematic viscosity and $\rho$ the density of the solvant. $\psi \vect{f}_p$ and $\vect{f}_{sq}$ are the forces necessary to enforce the particles rigidity and the required surface slip velocities $\vect{u}_{r}|_{r=a}$ and $\vect{u}_{\theta}|_{r=a}$ (given in the following paragraph). Finally, equation~\ref{NS_equation} is coupled to the underdamped Newton–-Euler equations for the center of mass and the angular velocity of the rigid particles. The self-propelling force thus originates from the interactions between the particles and the fluid, when we enforce the required slip velocities and assuming momentum conservation. Our simulations are nondimensionalized so that $\rho = \eta = \Delta =1$, $\Delta$ being the size of the integration grid. Rotational and translational thermal fluctuations can be added to these equations\cite{Iwashita08} even though most of the results shown here were obtained for $T=0$. Also, we will focus here on rather dilute regimes (with packing fractions $\phi$ generally close to 0.3, far from the jamming threshold) so that the continuous nature of the interface particle-fluid does not raise any controversy.\\

As for the particles, we are considering squirmers which means that we enforce the following ``squirming'' slip velocity conditions at the interface between the particles and the fluid:
\begin{equation}
\left\{
\begin{array}{ll}
\vect{u}_{r}|_{r=a} & = \sum_{n=1}^{\infty} A_n P_n\left( \cos{\theta}\right) \vect{\hat{r}} \\
& \\
\vect{u}_{\theta}|_{r=a} & = \sum_{n=1}^{\infty} \frac{2}{n(n+1)} B_n \sin{\theta}P'_n\left( \cos{\theta}\right) \vect{\hat{\theta}}
\end{array}
\right.
\label{Full_BC}
\end{equation}
with $\vect{\hat{r}}$ and $\vect{\hat{\theta}}$ the radial and tangential unit vectors, $\theta$ the angle of the orientation vector $\vect{e}$, $P_n$ the n-th order Legendre polynomial and $a$ the radius of the particles. In our case, we will consider the most simple squirmer with $A_n = 0$ for $\forall n$ and $B_n = 0$ for $\forall n > 2$. The slip-velocity is thus purely tangential:  
\begin{equation}
\vect{u}_{\theta}|_{r=a} = B_1 \sin{\theta} \left( 1 + \alpha \cos{\theta} \right) \vect{\hat{\theta}}
\end{equation}
where $\alpha = B_2 / B_1$ is the ratio between the amplitude of the first two squirming modes. It was shown analytically\cite{Ishikawa06, Zhu12} and numerically\cite{Molina13} that the swimming velocity of a squirmer is entirely determined by $B_1$ in Newtonian fluids. $\alpha$ on the other hand characterizes the self-propelling mechanism of the squirmer: if $\alpha > 0$, the particle is a ``puller'', if $\alpha < 0$ it is a ``pusher'' and if $\alpha = 0$, it is a ``neutral squirmer''. The velocity field generated by such swimmers can be calculated analytically\cite{Blake71} and the agreement between the numerical streamlines and the analytical ones is very good, thus validating our method\cite{SM}. In the bulk, the far-field velocity field generated by pushers and pullers is dominated by the Stokeslet dipole singularity and decays like $r^{-2}$ whereas it is dominated by the source dipole singularity for neutral squirmers and decays like $r^{-3}$ (see fig.~\ref{vfield} a)).

\begin{figure}[!ht]
\begin{tabular}{c}
\hspace{-0.5cm}
\includegraphics[width=3.8 cm]{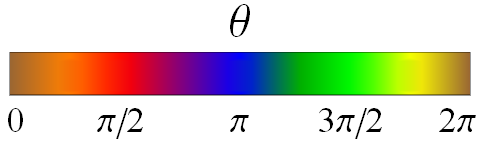} \hspace{0.5 cm}
\includegraphics[width=3.8 cm]{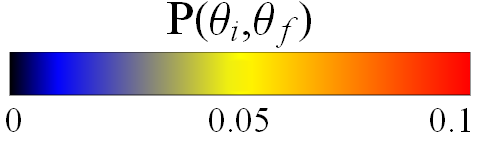}\\
\hspace{-0.5cm}
\includegraphics[width=3.6 cm]{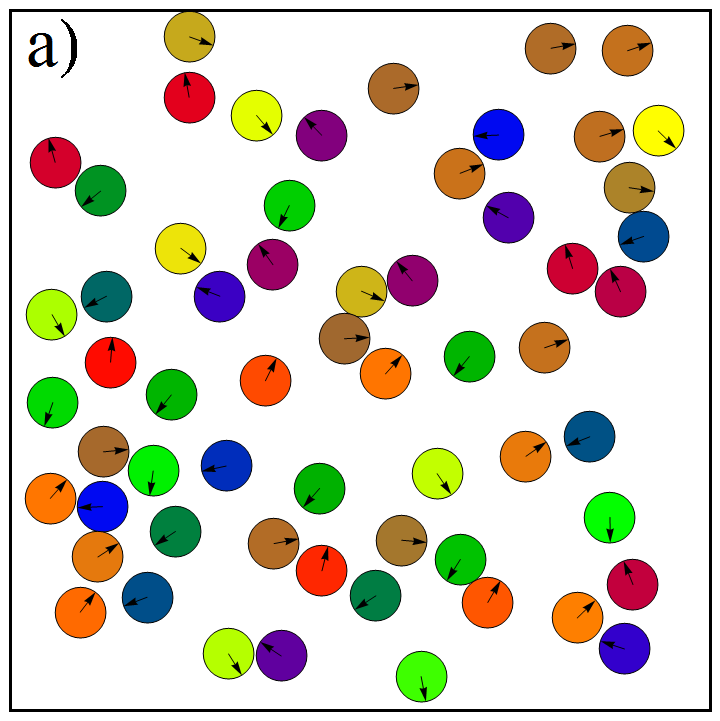} 
\includegraphics[width=4.4 cm]{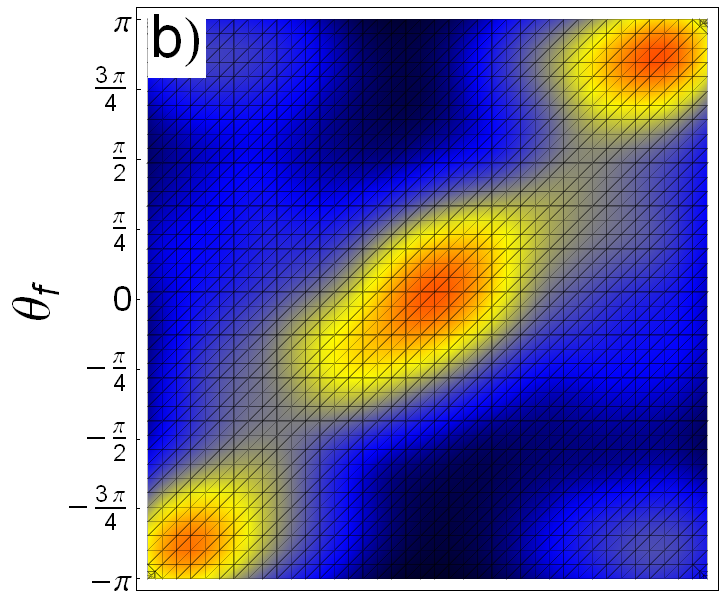}\\
\hspace{-0.5cm}
\includegraphics[width=3.6 cm]{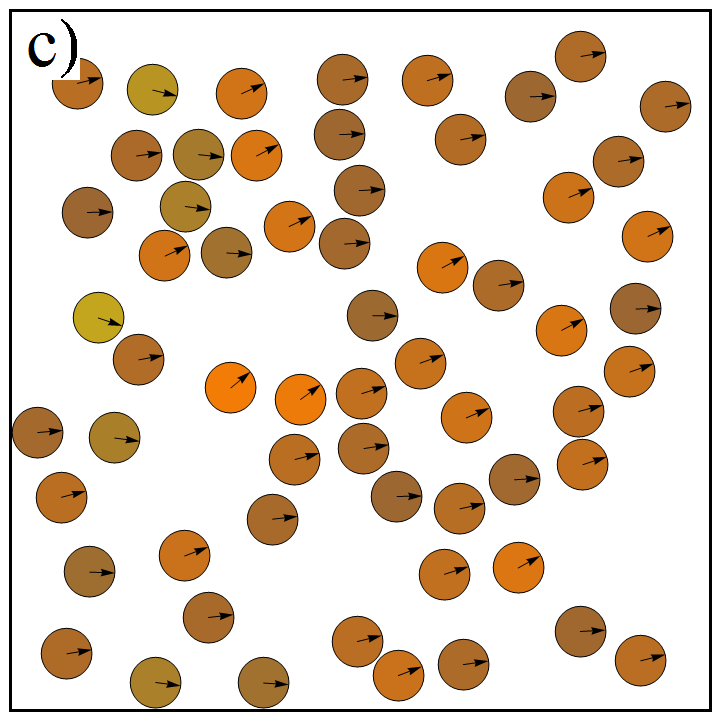} 
\includegraphics[width=4.4 cm]{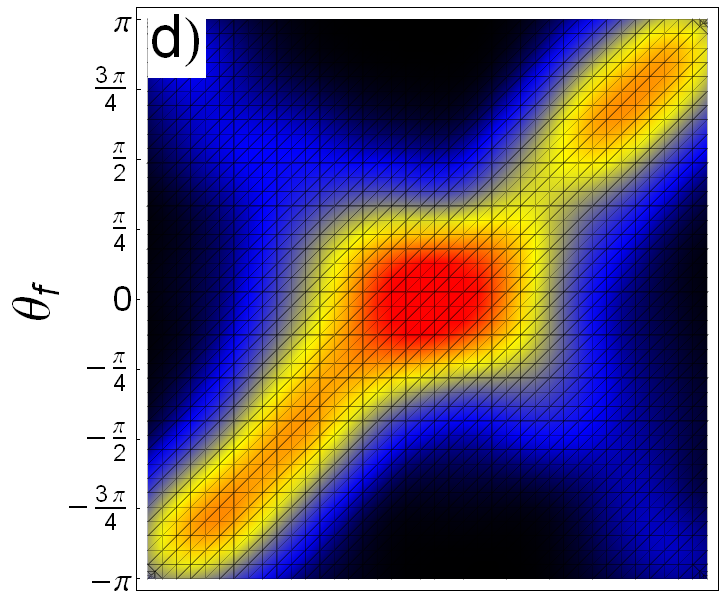} \\
\hspace{-0.5cm}
\includegraphics[width=3.6 cm]{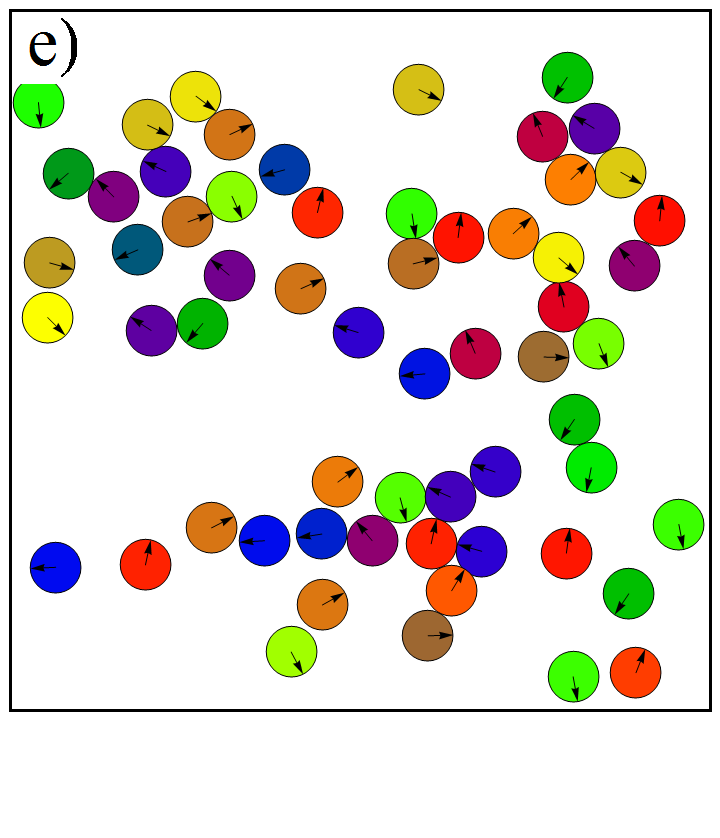} 
\includegraphics[width=4.4 cm]{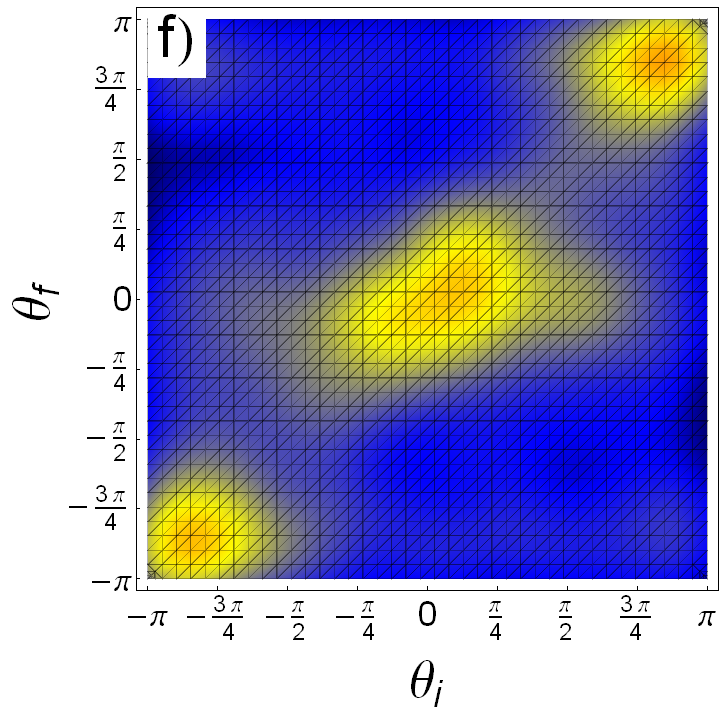} \\
\end{tabular}
\caption{\label{snapshots} Simulations of 60 squirmers starting from random positions and orientations at $T=0$. The packing fraction is $\phi \approx 0.29 $ and the Reynolds number $8 \times 10^{-3}$. a) and b) pushers with $\alpha=-5$, c) and d)  neutral squirmers, e) and f) pullers with $\alpha=5$. Left column: snapshots of the suspensions after 2 millions of time steps\cite{movies}. The arrows correspond to the orientation vectors of the particles and the colors to their angle $\theta$. Right column: probability density function $P(\theta_i, \theta_f)$ where $\theta_i$ and $\theta_f$ are the relative angles between two squirmers before and after they interact through near-field interactions respectively. We measured $\theta_i$ and $\theta_f$ when the distance between the two particles crosses a given threshold $ \| \vect{r}_{ij} \| = 2.6 a$. The colors correspond to the probability of the pair $\left(\theta_i, \theta_f \right)$ such that $\iint P\left(\theta_i, \theta_f \right) d\theta_i d\theta_f =1$. The angles were measured during the entire simulation (starting $t=0$).}
\end{figure}

The orientation vector of the squirmers $\vect{e}$ was kept two-dimensional here, with $\vect{e}_z(t) = 0$ for $\forall t$. This would correspond to experimental systems like Janus particles moving by self-induced electrophoresis, which are subjected to a torque preventing $\vect{e}$ to go out of the xy plane\cite{Suzuki11}. To prevent overlap, the particles also interact through a soft-core repulsive core potential (truncated Lennard-Jones). A lot of actual experiments on swimmers involve monolayers of particles or bacteria moving in quasi-two dimensional systems. To take into account the influence of the fluid confinement, we added to the simulations two walls with no-slip boundary conditions such as $u \left(z= \pm h/2 \right) = 0$ with $h$ the system height. In this paper, $h$ was kept constant at $h=5.2 \, a$. Under such a strong confinement, the Stokeslet dipole singularities are screened exponentially\cite{Liron76} and the source dipole becomes dominant in the far-field for all squirmers. Thus, the far-field flow decays like $r^{-2}$ independently of the self-propelling mechanism as can be seen on fig.~\ref{vfield} b) so that the hydrodynamic far-field interactions become generic.\\

Figure~\ref{snapshots} shows snapshots of suspensions of 60 squirmers with $\alpha = -5$, 0 and 5 after thousands of time steps, initially starting from random positions and orientations. One can immediately notice  the onset of directional motion in the system of neutral squirmers (fig.~\ref{snapshots} b): over time, particles align their orientation vectors, which is clearly shown on fig.~\ref{orderparam} by the regular increase of the system order parameter $ \langle e(t) \rangle = \| \sum_i^N \vect{e}_i(t) \| / N$ that eventually reaches 1. On the other hand, the suspensions of strong pullers or pushers shown on fig.~\ref{orderparam} never exhibit directional motion and their order parameter fluctuates around 0 at all times for high values of $|\alpha|$, even at zero temperature. The existence of globally aligned states in confined active suspensions has been predicted and observed numerically in simulations where the flow field is calculated using purely two-dimensional far-field approximations\cite{Brotto13,Lefauve14}. In that context, it was claimed that the distinction between the different types of squirmers should be irrelevant as they all generate similar far-field flows, as shown on figure~\ref{vfield} b). But in our simulations, the collective behavior of the suspension still clearly depends on the value of $\alpha$. We can draw two conclusions from here: first, the distinction between squirmers with different self-propelling mechanisms does not systematically become irrelevant under rigid confinement as only the neutral squirmers exhibit polarized states. Second, it seems that the collective behavior of an active suspension cannot always be predicted using only far-field approximation.\\

%In suspensions of neutral squirmers, the order parameter can still increase in the presence of translational and rotational diffusion (for $T \ne 0$). Moreover, we can see on the inset of figure~\ref{orderparam} that the packing fraction plays an important role too: suspensions of neutral squirmers with $ \phi \ge 0.34$ seem to be unable to exhibit directional motion as their order parameter barely increases during the whole simulation time. For computational reasons, we were not able yet to determine if it is fundamentally impossible or if takes longer but we can say for sure that increasing $\phi$ from $0.29$ to $0.34$ has a negative effect. On the opposite, increasing $\phi$ from $0.19$ to $0.24$ seems to favor this collective behavior as $ \langle e \rangle$ reaches 1 more quickly in the second case.\\

\begin{figure}[!ht]
\begin{tabular}{cc}
\includegraphics[width=8 cm]{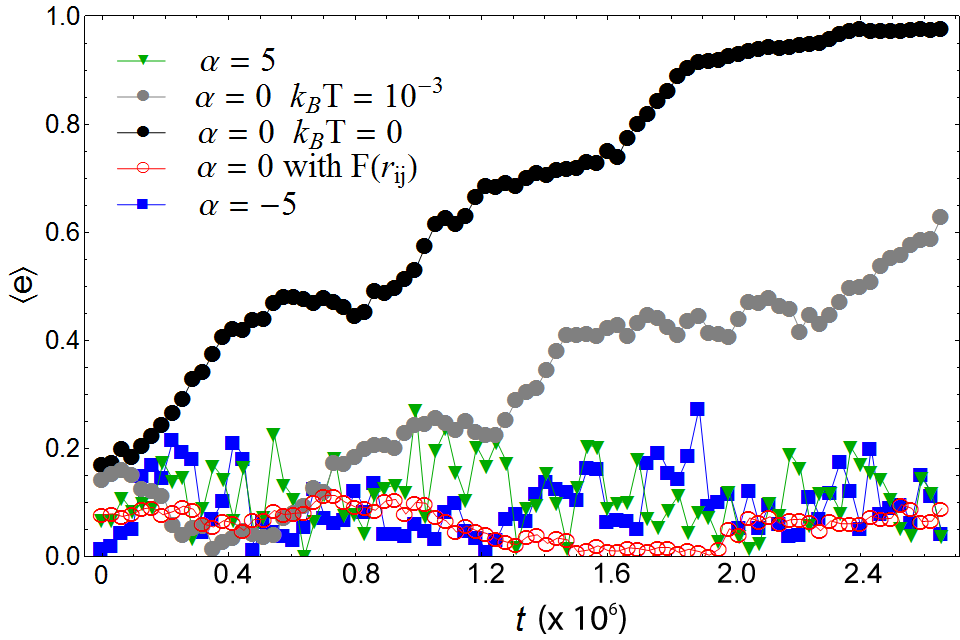}
\end{tabular}
\caption{\label{orderparam} Time evolution of the order parameter $\langle e \rangle$ in suspensions of 60 squirmers for different values of $\alpha$, temperatures and interaction forces. $t$ is the number of time steps. The squirmers start from random positions and orientations, the Reynolds number is $8 \times 10^{-3}$and the packing fraction is $\phi \approx 0.29 $. The gray circles and red empty circles show the influence of thermal fluctuations and of the additional repulsive force given by eq.~\ref{Fij} on the collective behavior in systems of neutral squirmers.}
\end{figure}

Our second conclusion suggests that the near-field hydrodynamic interactions also need to be taken into account. In order to understand their role, we ran simulations with an additional repulsive short-range force 
\begin{equation}
\vect{F}(r_{ij})  =
\left\{
\begin{array}{ll}
 \left[ \frac{U}{\lambda} \exp \left(\frac{-r_{ij}}{\lambda} \right) + c\right] \hat{\vect{r}}_{ij} & \mbox{if  } r_{ij} \le 4 a \\
   \vect{0} & \mbox{if  } r_{ij} > 4 a
\end{array}
\right.
\label{Fij}
\end{equation}
with $c$ a constant such that $F(4a)=0$ and $\lambda = 2a$ the characteristic length of the force. $U$ and $\lambda$ are tuned so that $\vect{F}(r_{ij})$ prevents the particles from being too close but remains small enough so that there is no jamming at this packing fraction. As $\vect{F}(r_{ij})$ has a cutoff equal to $4 a$, the far-field interactions between squirmers remain purely hydrodynamic and are exactly identical as in our previous simulations. Moreover, it has no direct effect on the relative alignment of the squirmers as it does not induce any torque. Figure~\ref{orderparam} shows that with this additional force, the order parameter $\langle e(t) \rangle$ fluctuates around zero during the whole simulation. Therefore without near-field hydrodynamic interactions, directional motion cannot be observed anymore which confirms that they play a crucial role in the collective behavior of confined active suspensions.\\

Ishikawa and Pedley\cite{Ishikawa06} found analytical expressions for the force and torque between squirmers in the near-field approximation (in the absence of confinement but here we assume the confinement will not modify the interactions at very small distances). The structure of the torque map appears quite different between pullers or pushers and neutral squirmers\cite{SM}. Moreover, the amplitude of the torque increases with $|\alpha|$. In order to see how this can influence the alignment between squirmers, we plotted the relative angle between the orientation vectors of two particles, before and after they interact at near-field distances (fig.~\ref{snapshots} right column). This plot can be viewed as some kind of ``collision-rules''\cite{Hanke13}, even if in our case a collision is not strictly defined and we have to choose an arbitrary distance to delimit this event. In the case of neutral squirmers, we can see that a lot of data points collapse on the straight line of slope 1 which means that many near-field interactions are ineffective with respect to the relative angle. However, one also has to note the presence of a red rectangle in the middle of the plot showing that for small $\theta_i$, we get almost all the time $|\theta_f| \le |\theta_i|$. This means that an effective alignment is induced by the near-field interactions at small angles. For pushers or pullers on the other hand, the near-field interactions have a much more complicated effect on the relative orientations and the same initial $\theta_i$ can lead to very different $\theta_f$, even at $T=0$. In particular, it is frequent to get $|\theta_f| \ge |\theta_i|$ even for low values of $\theta_i$, which corresponds to dis-aligning interactions. Note that this is particularly visible in the case of the pullers. One way to quantify the overall alignment induced by the near-field interactions is to compute the alignment integral defined as\cite{Hanke13}
\begin{equation}
\langle \Delta A \rangle  = |\cos{(\theta_i/2)}| - |\cos{(\theta_f/2)}|
\label{alignment}
\end{equation}
where the bracket $\langle \ldots \rangle$ is an average over all the possible collisions. Using the data of figure~\ref{snapshots} b), d) and f), we found $\langle \Delta A \rangle \approx 0.017$, $0.028$ and $0.005$ for $\alpha = -5$, $0$ and $5$ respectively. We thus have the confirmation that the effective alignment decreases when $|\alpha|$ increases.
\begin{figure}[!ht]
\begin{tabular}{cc}
\includegraphics[width=7 cm]{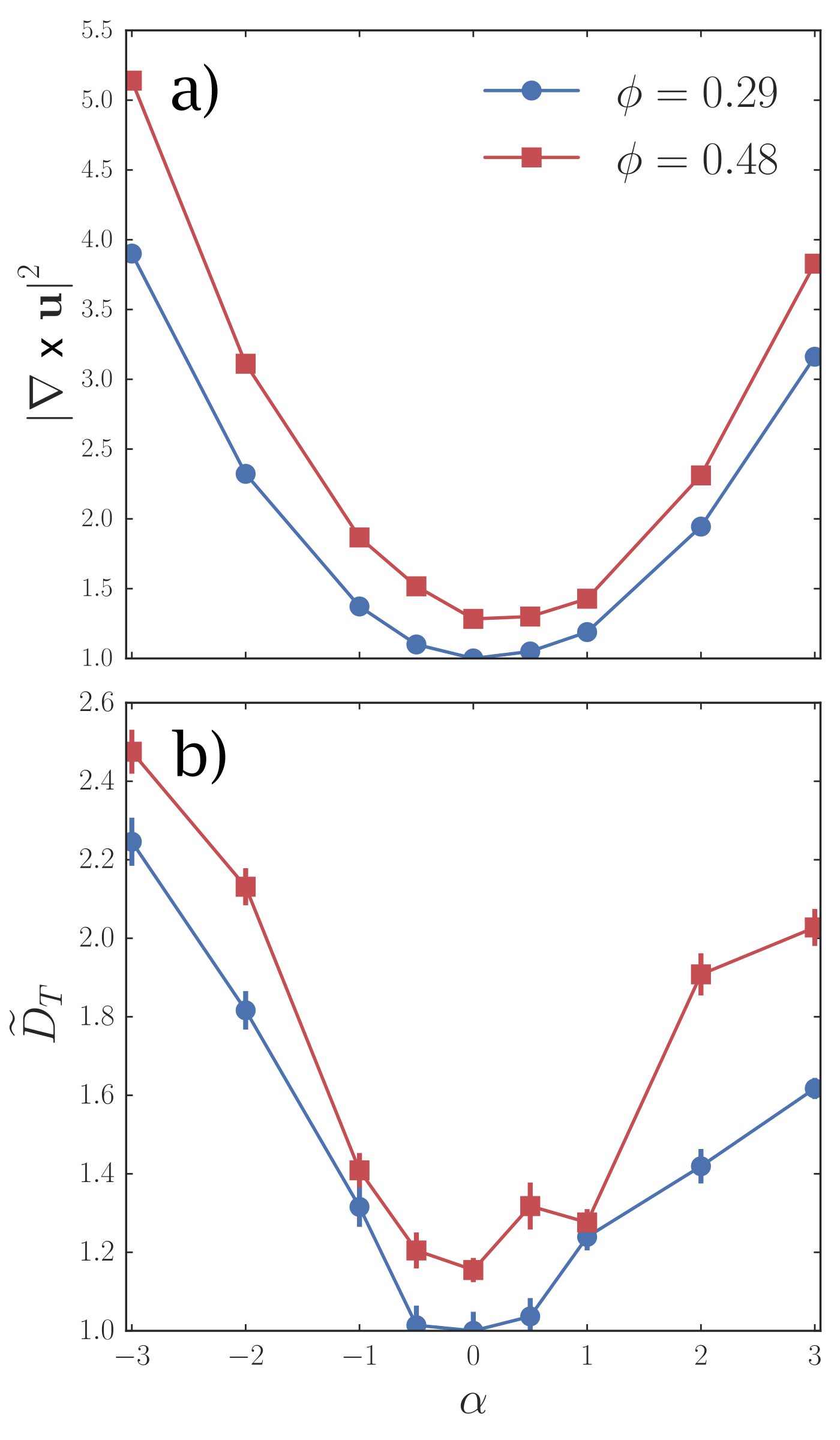}
\end{tabular}
\caption{\label{fluctuations}a) Enstrophy and b) effective diffusion coefficient with respect to $\alpha$ in suspensions of squirmers at T=0 for two different packing fractions. The values of $\tilde{D}_T$ are obtained by calculating the mean-square displacement of the particles. At long times, we have $\langle \Delta r^2 \rangle \propto \tilde{D}_T \, t$ so that we can extract the effective diffusion coefficients. The two plots have been normalized by $|\nabla \times u|^2(\alpha=0)=5.5 \times 10^{-6}$ and $\tilde{D}_T(\alpha=0) = 8.1 \times 10^{-2}$ respectively.}
\end{figure}

Another notable difference between the squirmers is the strength of the far-field flow they generate. As we can see on fig.~\ref{vfield} b), even if the far-field flow decays in the same way for all swimmers, $|\vect{u}(r)|$ still increases with $|\alpha|$. In order to quantify the disorder in hydrodynamic interactions for several suspensions, we calculated the averaged enstrophy $\langle  (\nabla \times \vect{u})^2 \rangle$ for several suspensions. Figure~\ref{fluctuations} a) shows that the enstrophy clearly increases with $|\alpha|$. It is then logical to assume that the fluid fluctuations should also increase with respect to this parameter, which is confirmed by the numerical measurements of the effective diffusion coefficient of the particles\cite{Molina13} shown on fig.~\ref{fluctuations} b). As a consequence, increasing $|\alpha|$ is equivalent to increasing the strength of the orientational noise which is known to prevent the onset of directional motion. Note that a systematic asymmetry between pullers and pushers can once again be observed. \\

Recent numerical and theoretical studies have proved that the onset of directional motion strongly depends on how much particles align their orientations during binary collisions. Hanke and collaborators showed that collision rules generating a weak relative alignment were enough to induce polarized state for similar packing fractions at low noise\cite{Hanke13}. Simulations of active hard-disks highlighted that the details of the scattering map do not matter as long as an effective alignment takes place for low angular separation collisions\cite{Lam15}. Figure~\ref{snapshots} d) clearly shows that there is such a relative alignment at small angles for neutral swimmers. When $|\alpha|$ increases however, the alignment becomes weaker as shown by the decrease of the alignment integral $\langle \Delta A \rangle$ and the strength of the orientational noise increases simultaneously as shown by the increase of the effective diffusion coefficient. We believe that these two phenomena explain why directional motion cannot be observed in suspensions of strong pushers or pullers. Even if we report here simulations of small systems, we believe the eventual finite size effects will not change qualitatively these conclusions as the suspensions are big enough to see the influence of the confinement on the far-field hydrodynamic flow field (see fig.~\ref{vfield} b)).\\

In summary, we have shown that in confined geometries, the collective behavior of an active suspension still strongly depends on the self-propelling mechanism of the swimmers. Despite the fact that pushers, neutral squirmers and pullers generate similar far-field hydrodynamic fields in thin suspensions, stable polar states were only observed for small values of $\alpha$. Our results suggest that collective behaviors cannot be predicted only by using far-field approximations of the flow field. In particular, one also has to look at the scattering map induced by the combination of near-field hydrodynamic forces and steric interactions. For squirmers, we demonstrated that an effective alignment exist for collisions at small relative angles. When the parameter $|\alpha|$ increases, the reduction of this effective alignment coupled to the increase of the amplitude of the fluid fluctuations make the polar states unstable. We believe our results help us understand why there are so few experimental observations of directional motion in active suspensions: even if these ones are strongly confined, the formation of polar states is still strongly dependent on the nature of the mutual near-field interactions, which will be function of the self-propelling mechanism of the swimmers. To our knowledge, many artificial swimmers used in experiments like Janus particles can be considered as pushers or pullers, for which we have seen that this effective alignment was quite weak. 

%**************acknowledgments*************

\renewcommand{\abstractname}{Acknowledgements}
\begin{abstract}
 The authors wish to thank Professor Ryoichi Yamamoto for helpful discussions. This work is supported by the Japan Society for the Promotion of Science Grants-in-Aid for Scientific Research KAKENHI. 
\end{abstract}

\bibliographystyle{siam}
\bibliography{biblio}

\newpage
\onecolumn

{\huge Supplementary material: collective behavior of strongly confined suspensions of squirmers}
\vspace{1cm}
\section{Comparison between numerical and analytical streamlines}

\begin{figure}[!ht]
\centering
\begin{tabular}{ccc}
\includegraphics[width=4 cm]{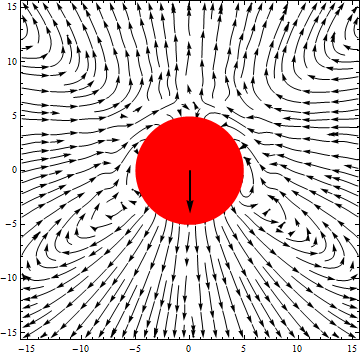} &
\includegraphics[width=4 cm]{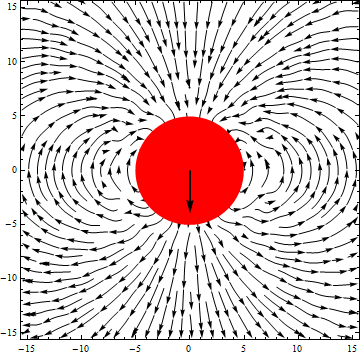} &
\includegraphics[width=4 cm]{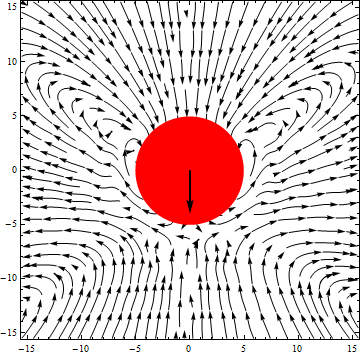} \\
\includegraphics[width=4 cm]{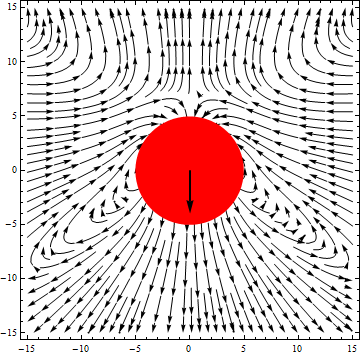} &
\includegraphics[width=4 cm]{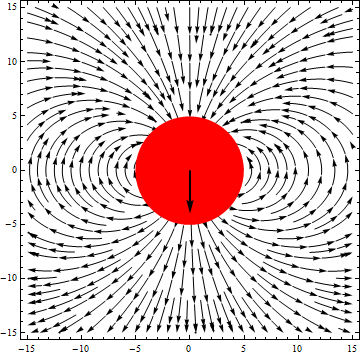} &
\includegraphics[width=4 cm]{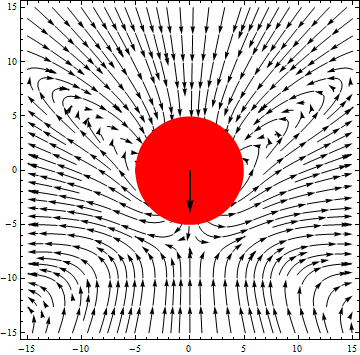} 
\end{tabular}
\caption{\label{streamlines} Comparison between the numerical and analytical streamlines induced by the three different kinds of squirmers. Top row: numerical streamlines. Bottom rows: analytical streamlines given by equation~\ref{Blake}. Left column: pusher with $\alpha = -1$. Middle column: neutral squirmer. Right column: puller with $\alpha = 1$. The black arrow shows the orientation of the squirmer.}
\end{figure}

One very big advantage of the model of squirmers is that it is possible to find analytical expressions for the velocity and pressure fields induced by the motion of these particles. Solving the Stokes equation with the boundary conditions of squirmers given by equation (2), Blake found that\cite{Blake71}:

\begin{equation}
\begin{array}{ll} 
\boldsymbol{u}(\boldsymbol{r})  & = \frac{a^3}{r^3} B_1 \left[ \left( \boldsymbol{e}.\boldsymbol{\hat{r}} 
\, \boldsymbol{\hat{r}} - \frac{1}{3} \boldsymbol{e} \right)
+ \left(\frac{a^{4}}{r^{4}} -\frac{a^{2}}{r^{2}} 
\right) \alpha P_2 \left( \boldsymbol{e}.\boldsymbol{\hat{r}} \right)\,
\boldsymbol{\hat{r}} + \frac{a^{4}}{r^{4}} \alpha W_2 \left(\boldsymbol{e}.\boldsymbol{\hat{r}} \right)
\left(\boldsymbol{e}.\boldsymbol{\hat{r}} \, 
\boldsymbol{\hat{r}} - \boldsymbol{e} \right) \right]
\end{array}
\label{Blake}
\tag{S1}
\end{equation}
where the function $W_n$ is defined from the Legendre polynomial $P_n$ as
\begin{equation}
W_n(\boldsymbol{e}.\boldsymbol{\hat{r}}) = \frac{2}{n(n+1)} P_n' \left( \boldsymbol{e}.\boldsymbol{\hat{r}} \right)
\tag{S2}
\end{equation}
To validate our numerical method, we have compared the numerical streamlines induced by a single particle in the bulk to equation~\ref{Blake}. As can be seen on figure~\ref{streamlines}, the agreement is excellent.

\newpage

\section{Hydrodynamic torque between squirmers}

\begin{figure}[!ht]
\centering
\begin{tabular}{ccc}
\includegraphics[width=4 cm]{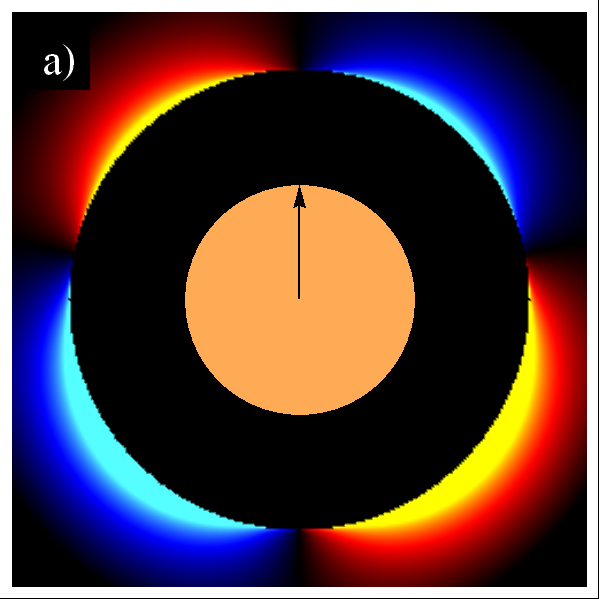} &
\includegraphics[width=4 cm]{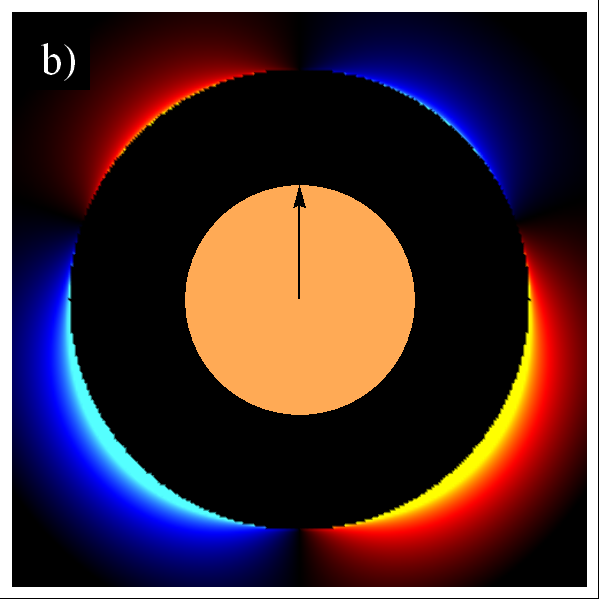} &
\includegraphics[width=4 cm]{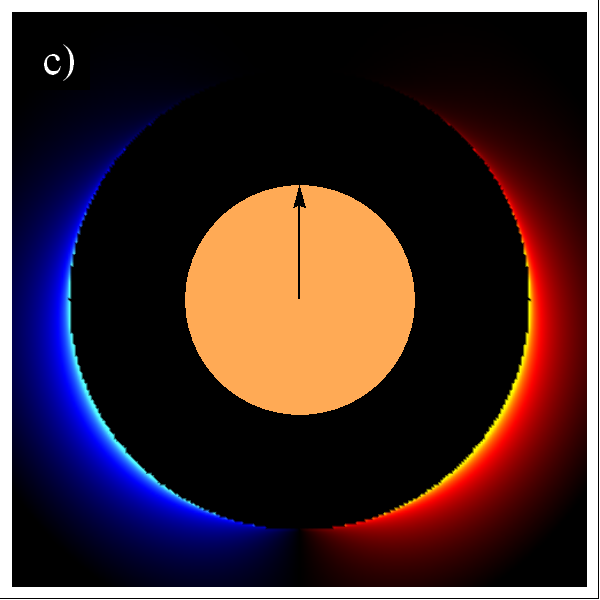} \\
\includegraphics[width=2 cm]{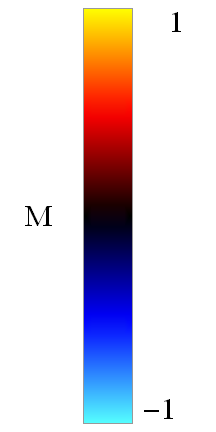} &
\includegraphics[width=4 cm]{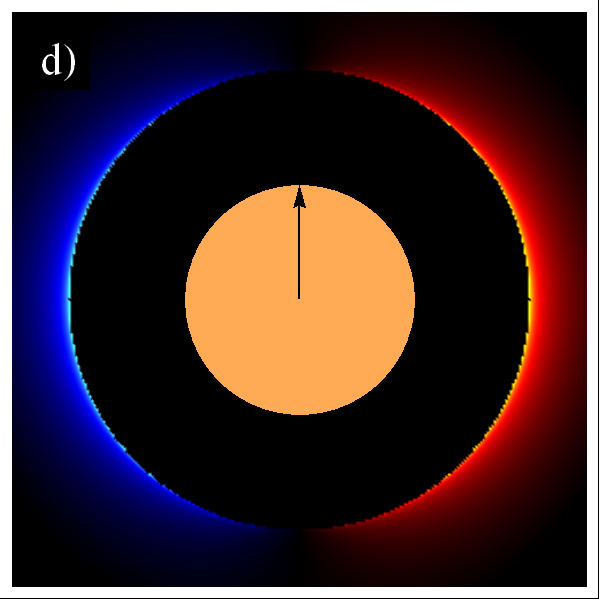} & \\
\includegraphics[width=4 cm]{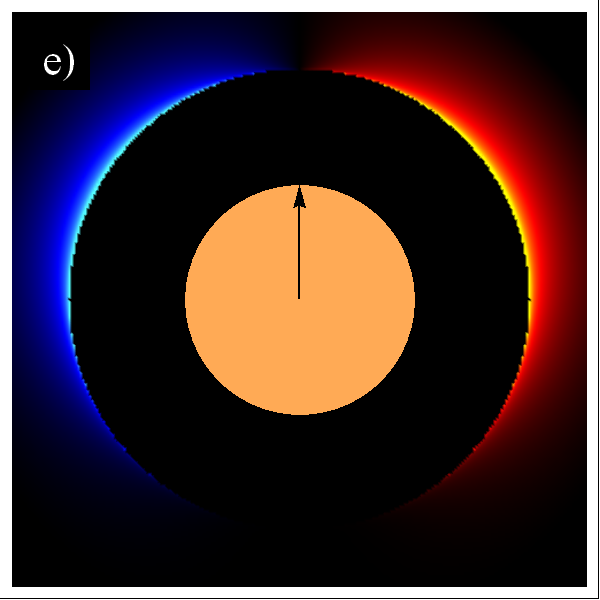} &
\includegraphics[width=4 cm]{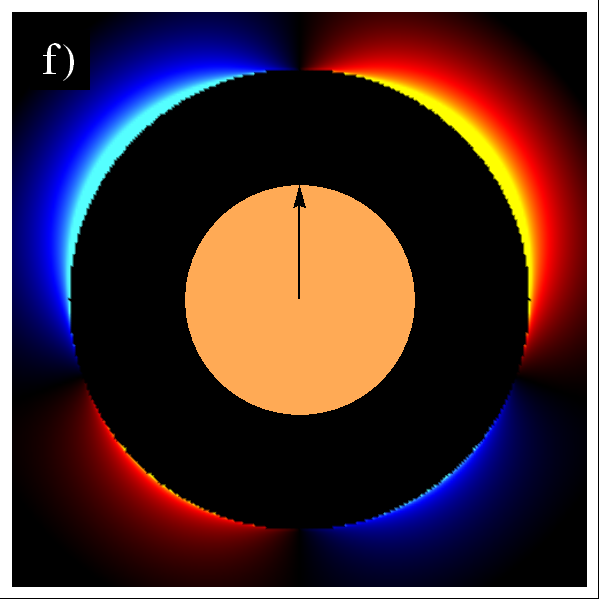} &
\includegraphics[width=4 cm]{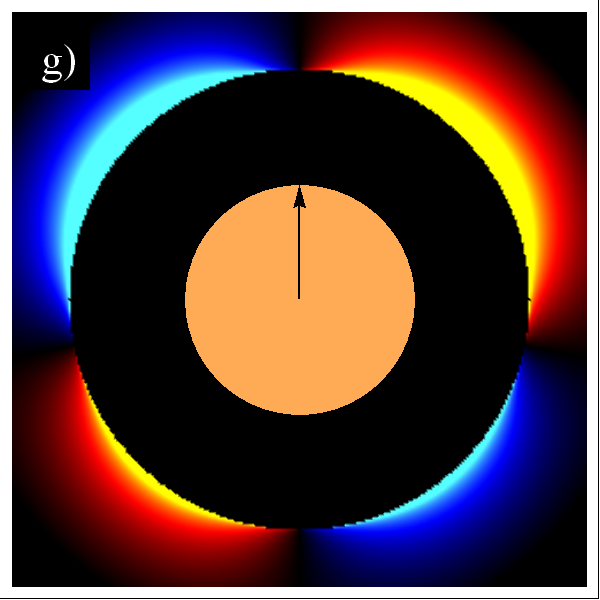}
\end{tabular}
\caption{\label{torque} z-component $T_z$ of the torque induced by a squirmer according to equation~\ref{torque_Ishikawa} in arbitrary units. From top-left to bottom right, $\alpha = -5$, $-3$, $-1$, $0$, $1$, $3$ and $5$ respectively. The orange disk in the middle of each plot corresponds to the squirmer and the arrow to its orientation vector $\boldsymbol{e}$. The black disk surrounding it comes from the fact that steric interactions prevents particles from overlapping, so that $r_{ij} \ge 2 \, a$.}
\end{figure}

Using Blake's results, Ishikawa and Pedley found analytical expressions for the torque induced by a squirmer in the near-field approximation\cite{Ishikawa06}. Its z-component (the only one not equal to zero in our case) is given in the first approximation by:
\begin{equation}
T_z\left( \boldsymbol{r_{ij}} \right)  = - \mu \pi a^2 \boldsymbol{e}.\boldsymbol{\hat{x}_{ij}} \, B_1 \left[ W_1\left(-\boldsymbol{e}.\boldsymbol{\hat{y}_{ij}}\right) + \alpha W_2\left(-\boldsymbol{e}.\boldsymbol{\hat{y}_{ij}}\right)\right]  \log(\boldsymbol{r_{ij}}^2) 
\tag{S3}\label{torque_Ishikawa}
\end{equation}

with $\mu$ the viscosity of the fluid. Using this expression, we plotted the torque map for different values of $\alpha$ (figure~\ref{torque}). Whereas it is dipolar for $\alpha = 0$, it gets more and more quadrupolar as $|\alpha|$ increases, which was confirmed recently by numerical works~\cite{Molina14}. Note that the intensity of the torque increases when $|\alpha|$ increases. The zones of strong torques correspond to the recirculation zones of the fluid. The near-field hydrodynamic interactions between strong pullers or pushers and neutral squirmers being particularly different, it is not surprising that their influence on the relative orientations of the particles depends on the value of $\alpha$.

\end{document}